\documentclass[twocolumn,aps,prl,showpacs,preprintnumbers,floatfix,nofootinbib,superscriptaddress]{revtex4}
\usepackage{epsfig}
\usepackage{amsmath,amssymb}
\newcommand{\be}{\begin{eqnarray}}
\newcommand{\ee}{\end{eqnarray}}

\newcommand{\lp}{\ell_{\rm P}}
\newcommand{\mpl}{M_{\rm P}}
\renewcommand{\lg}{\ell_{\rm G}}
\newcommand{\mg}{M_{\rm G}}
\newcommand{\rh}{R_{\rm H}}
\newcommand{\lc}{\lambda_{\rm C}}

\newcommand{\sen}{\tilde S^{\rm E}_{(4)}}

\newcommand{\segn}{\tilde S^{\rm E}_{(4+d)}}

\newcommand{\sefn}{\tilde S^{\rm E}_{\rm eff}}
\newcommand{\meff}{M_{\rm deg}}

\newcommand{\mc}{M_{\rm C}}
\newcommand{\sesub}{\tilde S^{\rm E}_{\rm sub}}
\begin{document}
\title{Brane-world black holes and the scale of gravity}
\author{G.L.~Alberghi}
\email{alberghi@bo.infn.it}
\affiliation{Dipartimento di Fisica, Universit\`a di Bologna, via Irnerio~46, 40126 Bologna Italy}
\affiliation{Istituto Nazionale di Fisica Nucleare, Sezione di Bologna}
\author{R.~Casadio}
\email{casadio@bo.infn.it}
\affiliation{Dipartimento di Fisica, Universit\`a di Bologna, via Irnerio~46, 40126 Bologna Italy}
\affiliation{Istituto Nazionale di Fisica Nucleare, Sezione di Bologna}
\author{O.~Micu}
\email{micu.octavian@gmail.com}
\affiliation{Lessingstr.~19, Dortmund D-44147, Germany}
\author{A.~Orlandi}
\email{orlandi@bo.infn.it}
\affiliation{Dipartimento di Fisica, Universit\`a di Bologna, via Irnerio~46, 40126 Bologna Italy}
\affiliation{Istituto Nazionale di Fisica Nucleare, Sezione di Bologna}
%
%
%
%
\begin{abstract}
A particle in four dimensions should behave like a classical black hole  
if the horizon radius is larger than the Compton wavelength or, equivalently,
if its degeneracy (measured by entropy in units of the Planck scale) is large.
For spherically symmetric black holes in $4+d$ dimensions, both arguments again
lead to a mass threshold $\mc$ and degeneracy scale $\meff$ of the order of the
fundamental scale of gravity $\mg$.
In the brane-world, deviations from the Schwarzschild metric induced by bulk
effects alter the horizon radius and effective four-dimensional Euclidean action
in such a way that $\mc\simeq\meff$ might be either larger or smaller than $\mg$.
This opens up the possibility that black holes exist with a mass smaller
than $\mg$ and might be produced at the LHC even if
$\mg\gtrsim 10\,$TeV, whereas effects due to bulk graviton exchanges remain
undetectable because suppressed by inverse powers of $\mg$.
Conversely, even if black holes are not found at the LHC, it is still possible that
$\mc\gg\mg$ and $\mg\simeq 1\,$TeV.
\end{abstract}
\pacs{}
\maketitle
\noindent
{\underline{\em Introduction\/}}
A most exciting feature of string-inspired models with large extra
dimensions~\cite{arkani,RS} is that the fundamental scale of gravity
$\mg$ could be much smaller than the Planck mass $\mpl\simeq 10^{16}\,$TeV
and as low as the electro-weak scale ($\mg\simeq 1\,$TeV).
Microscopic black holes may therefore be created in our
accelerators~\cite{banks,dimopoulos,cavaglia}
with a production cross section given, according to the hoop conjecture~\cite{thorne}, 
by $\sigma\sim \rh^2$~\cite{dimopoulos}, where $\rh$ is the radius of the
forming horizon and is bounded below by the wavelength of typical quantum
fluctuations~\cite{hsu,dvali}.
After the black hole has formed (and possible transients),
the Hawking radiation~\cite{hawking} is expected to set off,
with the most common description based on the
canonical Planckian distribution for the emitted particles and
consequent instantaneous decay~\cite{dimopoulos}.
This standard picture and a variety of refinements have been implemented
in the most recent Monte~Carlo codes~\cite{MC}
and the outcome is being confronted with Large~Hadron~Collider
(LHC)~data~\cite{cms,atlas}.
One problem with the canonical description is that the black hole specific
heat is in general negative and one should therefore use the more consistent
microcanonical description~\cite{mfd,bcLHC},
which however requires an explicit counting
of the black hole microscopic degrees of freedom (or degeneracy).
\par
For this counting, one may appeal to the {\em area law\/}~\cite{hawk1971},
from which it can be inferred the horizon area describes the black
hole degeneracy~\cite{bek,hl}.
The area-entropy correspondence has inspired the holographic
principle~\cite{[7]} in order to solve the black hole information
paradox~\cite{maldacena}.
This principle has widely been developed~\cite{[8]}, and a 
theoretical support was found in the AdS/CFT correspondence~\cite{[10]}
that conjectures the equivalence of a string theory with gravity
in anti-de~Sitter space with a quantum field theory
without gravity on the boundary.
In this letter we shall analyze the interplay between the classicality condition
that must be met in black hole formation and the horizon area as a measure
of the entropy of black holes in the brane-world~\cite{RS}.
Our results allow for the existence of ``lightweight'' microscopic
black holes (LBH) with mass below $\mg$, or could explain the non existence
of black holes within the reach of LHC experiments~\cite{cms,atlas} even if
$\mg\simeq 1\,$TeV.
\par
\noindent
{\underline{\em Compton classicality\/}}
A black hole is a classical space-time configuration and its production
in a collider is therefore a ``classicalization'' process in which quantum
mechanical particles are trapped by gravitational self-interaction
within the horizon~\cite{hsu,dvali}.
Consequently, quantum fluctuations should be negligible for the final
state (of total energy $M$) which sources such a metric.
A widely accepted condition of classicality is then expressed
by assuming the Compton wavelength
$\lc\simeq\hbar/M=\lp\,\mpl/M$~\footnote{We shall mostly use units with the
Boltzmann constant $k_{\rm B}=c=1$, $G_{\rm N}=\lp/\mpl$
and $\hbar=\lp\,\mpl=\lg\,\mg$.}
of the black hole, viewed as {\em one particle\/}, is the lower bound for the
``would-be horizon radius'' $\rh$, that is
\be
\rh\gtrsim \lc
\ ,
\label{Ccond}
\ee 
where $\rh=\rh(M)$ depends on the specific black hole metric.
In four dimensions, using the Schwarzschild metric, one
obtains
\be
\rh=2\,\lp\,\frac{M}{\mpl}
\gtrsim
\lp\,\frac{\mpl}{M}
\quad
\Rightarrow
\quad
M\gtrsim\mc\simeq \mpl
\ ,
\label{condC}
\ee
which is supported by perturbative calculations of scattering amplitudes
for particles with centre-mass energy $M$~\cite{dvali}.
The above derivation does not make full use of the space-time geometry and, 
in particular, neglects that, for $M$ approaching the scale $\mg$, quantum fields
should be affected by extra-spatial dimensions (if they exist).
\par
\noindent
{\underline{\em Entropic classicality\/}}
Another classicality argument can be given, which does not involve black hole
wavefunctions but relies on Bekenstein's conjectured correspondence between
the entropy of thermodynamical systems and the area of black hole horizons~\cite{bek}.
Christoudolou~\cite{chris} first pointed out that the irreducible mass $M_{\rm ir}$
of a Kerr black hole, {\em i.e.}~the amount of energy that cannot be converted into work
by means of the Penrose process~\cite{penrose}, is related to the horizon area $A$
as $M_{\rm ir}=\mpl\,\sqrt{A/16\,\pi\,\lp^2}\equiv\sqrt{A_{\rm ir}}$.
Now, in thermodynamics, an increase in entropy is associated with
a degradation of energy because the work we can extract from the system
is reduced.
The similarity is clear, but goes beyond this simple statement.
For a Schwarzschild black hole, $M_{\rm ir}=M$ and no energy at all can be extracted.
Nonetheless, we can take a collection of fully degraded
subsystems (Schwarzschild black holes) and still get some work out of them.
In fact, if we merge two or more black holes, the total horizon area must equal
at least the sum of all their original areas~\cite{hawk1971}.
Denoting by $M_i$ and $A_i$ the initial irreducible masses and areas,
and by $M_F$ and $A_F$ the final irreducible mass and horizon area,
we see that
$M_F=\sqrt{A_F} = \sqrt{\sum_i A_i} < \sum_i \sqrt{A_i}= \sum_i M_i$.
The final irreducible mass is then less than the sum of all the initial irreducible
masses:
some more work can be extracted by merging fully degraded black holes.
The same occurs by collecting thermodynamical systems that -- individually --
are fully degraded but together can still provide work.
Bekenstein~\cite{bek} remarked how we can clarify these similarities
by invoking Shannon's entropy
\be
S = -\sum_n\, p_n\, \ln p_n
\ ,
\ee
where $p_n$ is the probability for a thermodynamical system to be found
in the $n$-th state.
A thermodynamical system is described in terms of a few macroscopical
variables (like energy, temperature and pressure).
Once these variables are fixed, the system can however be described
by a huge amount of \emph{microscopically inequivalent\/} states.
Hence, entropy can be seen as the lack of information about
the actual internal structure of the system.
Analogously, any four-dimensional black hole can be described in terms of
three macroscopic variables:
mass, angular momentum and charge.
All information about the matter which formed the black hole is
lost beyond the horizon.
Because of properties shared by thermodynamical entropy
and horizon area, Bekenstein found the simplest expression
(with dimensions of $\hbar$)
which satisfies the conditions on the irreducible mass is
\be
S_{\rm BH}
=
\frac{\mpl\,A}{16\,\pi\,\lp}
\ .
\label{Sbh}
\ee
Using a \emph{gedanken experiment\/}, Bekenstein \cite{bekbound} further
obtained the so-called entropy bound $S_{\rm BH} \leq 2\, \pi\, \rh\, M$,
and this topic has by now been extended to more general scenarios
(for a review, see Ref.~\cite{bousso}).
%
%
\par
From Eq.~\eqref{Sbh}, we can now infer an {\em entropic condition\/}
for black hole classicality:
a four-dimensional classical black hole should have a large degeneracy
(in units of the Planck scale), that is
\be
\sen
\equiv
\frac{S_{\rm BH}}{\lp\,\mpl}
=
\frac{4\,\pi\,\rh^2}{16\,\pi\,\lp^2}
\simeq
\left(\frac{M}{\meff}\right)^2
\gtrsim 1
\ ,
\label{condB}
\ee
which, for the Schwarzschild metric, leads to $M\gtrsim\meff\simeq\mpl$.
This conclusion is also supported by perturbative calculations of scattering amplitudes,
since the entropy~\eqref{Sbh} can be reproduced by assuming
the final classical black holes are composed of quanta with wavelength
$\lambda\sim\rh$~\cite{dvali}.
Note, however, that the physical meaning of the two scales 
is not quite the same: 
$\meff$ is the natural unit for measuring black hole internal degrees
of freedom (like the gap between energy levels of the harmonic oscillator),
whereas $\mc$ is the minimum mass below which black holes do not exist
(like the threshold in massive particle production).
That $\meff\simeq\mc\simeq\mpl$ is expected -- because gravity
in four dimensions entails one scale -- but is till a remarkable evidence
that black holes hide most information about forming matter.
\par
\noindent
{\underline{\em ADD black holes\/}}
Both classicality conditions~\eqref{condC} and~\eqref{condB} can be straightforwardly
generalized to models with extra-spatial dimensions by replacing $\mpl$ and
$\lp$ with $\mg$ and $\lg$, and using the appropriate expressions
for the horizon radius.
For example, in the ADD scenario of Refs.~\cite{arkani}, the brane tension is neglected
and one can therefore consider vacuum solutions to the Einstein equations
in $4+d$ dimensions to derive the following relation between the mass and
horizon radius~\cite{dimopoulos},
\be
R_{\rm H}=\frac{\lg}{\sqrt{\pi}}\,
\left(\frac{M}{M_{\rm G}}\right)^{\frac{1}{1+d}}
\left(\frac{8\,\Gamma\!\left(\frac{d+3}{2}\right)}{2+d}
\right)^{\frac{1}{1+d}}
\ ,
\ee
where $\Gamma$ is the usual Gamma function.
Inserting the above into Eq.~\eqref{Ccond} yields~\footnote{This is the 
kind of condition employed in all Monte Carlo studies of black hole production
at the LHC~\cite{MC}.}
\be
\rh
\gtrsim
\lg\,\frac{\mg}{M}
\qquad
\Rightarrow
\qquad
M\gtrsim \mc\simeq \mg
\ ,
\label{M>Mg}
\ee
as one would naively expect.
Moreover, the same result is again obtained by generalizing the
entropic argument to $4+d$ dimensions, namely
\be
\segn
\simeq
\left(\frac{\rh}{\lg}\right)^{2+d}
\sim
\left(\frac{M}{\meff}\right)^{\frac{2+d}{1+d}}
\gtrsim
1
\ ,
\ee
where $\meff\simeq\mg$.
One therefore concludes that even in the ADD scenario, gravity enters
black hole physics with one scale, $\mg$, like in four dimensions.
\par
\noindent
{\underline{\em Brane-world black holes\/}}
The situation appears more involved in the brane-world (RS)
scenario~\cite{RS}, in which the brane tension is not ignored
and the bulk is consequently warped.
This has made it very hard to describe black
holes~\cite{bwbh}~\footnote{Arguments have been formulated against
the existence of static brane-world black hole metrics~\cite{stability}.
Given the Hawking radiation is likely a strong effect for microscopic
black holes, their instability is here taken as granted.},
and only a few analytical candidates are known which solve the effective
four-dimensional vacuum Einstein equations~\cite{shiromizu},
\be
R_{\mu\nu}-\frac{1}{2}\,R\,g_{\mu\nu}
=
\mathcal{E}_{\mu\nu}
\qquad
\Rightarrow
\qquad
R=0
\ ,
\ee
where the presence of tidal effects from the propagation of gravity
into the bulk is represented by the (traceless) projected Weyl tensor
$\mathcal{E}_{\mu\nu}$.
One of these solutions is the tidally charged metric~\cite{dadhich} 
\be
ds^2=-A\,dt^2+A^{-1}\,dr^2+r^2\left(d\theta^2+\sin^2\theta\,d\phi^2\right)
\ ,
\label{tidalg}
\ee
with
\be
A=1-\frac{2\,\lp\,M}{\mpl\,r}-q\,\frac{\lg^2}{r^2}
\ ,
\ee
which has been extensively studied in Refs.~\cite{bcLHC,bhEarth,cfm,bulk}.
In the above and what follows, the tidal charge $q$ and $M$ are treated as
independent quantities, although one expects $q$ vanishes when the black
hole mass $M=0$.
Further, we only consider the case $q>0$ since negative tidal charge
would yield anti-gravity effects~\cite{cfm}.
A relation $q=q(M)$ should be obtained by solving the complete five-dimensional
Einstein equations~\cite{bulk} (or by means of supplementary arguments~\cite{ovalle}).
Nevertheless, since we are here interested in black holes near their minimum possible
mass $\mc\sim\mg\ll\mpl$,
we can approximate $q\simeq q(\mc)$ and constant, and expand all final expressions
for $M\sim\mg\ll\mpl$.
\par
We can first apply the usual classicality argument~\eqref{Ccond},
with the horizon radius
\be
\rh=\lp\left(
\frac{M}{\mpl}
+\sqrt{\frac{M^2}{\mpl^2}+q\,\frac{\mpl^2}{\mg^2}}
\right)
\ ,
\ee
and obtain $M\gtrsim \mc$, where the minimum mass
\be
\mc
\simeq 
\frac{\mg}{\sqrt{q}}
\ ,
\label{min}
\ee
for $M\sim\mg\ll\mpl$.
\par
We can also repeat the entropic argument by employing the effective
four-dimensional action, namely
\be
\sefn
\simeq
\frac{4\,\pi\,\rh^2}{16\,\pi\,\lp^2}
\ ,
\label{Se}
\ee
which, however, is not at a minimum for $M\simeq\mc$, as one
would instead expect from previous cases.
This discrepancy can be cured by recalling the Euclidean action
(as well as the thermodynamical entropy)
is defined modulo constant terms, which, for example, do not affect 
the value of the Hawking temperature nor the microcanonical description
of the Hawking radiation~\cite{mfd}.
By subtracting from Eq.~\eqref{Se} a suitable constant, namely
\be
\sesub=\sefn(M)-\sefn(\mc)
\ ,
\ee
and expanding for $M\sim\mg\ll\mpl$, we finally obtain
\be
\sesub
\simeq
\frac{M}{\meff}
\ ,
\ee
where there now appears the effective degeneracy scale
\be
\meff
\simeq 
\frac{\mg}{\sqrt{q}}
\ .
\ee
It is again remarkable that $\meff\simeq\mc$ and brane-world black holes are also
described by one scale [recall that $q\simeq q(\mc)$ is not truly independent]. 
The ``natural'' choice would now be $q\simeq 1$, so that $\meff\simeq\mc\simeq\mg$,
but the effective scale $\meff\simeq\mc$ could also be either larger ($q\ll 1$)
or smaller ($q\gg 1$) than $\mg$.
\par
\noindent
{\underline{\em Concluding remarks\/}}
Detection of black holes at the LHC would be a clear signal that we are embedded in a
higher-dimensional space-time and the fundamental scale of gravity $\mg\simeq 1\,$TeV.
The existence of extra spatial dimensions could also be uncovered by means of particle
processes which involve the exchange of bulk gravitons~\cite{giudice}.
Such processes are perturbatively described by operators suppressed
by inverse powers of $\mg$, and might not be detectable if the latter
is larger than a few TeV~\cite{strumia}.
We have shown that both classicality conditions, from quantum mechanics 
and the entropic counting of internal degrees of freedom, allow for brane-world
black holes with minimum mass~\eqref{min}.
The latter might be different from the fundamental scale $\mg$, if $q$ departs
significantly from $1$ (which should be related to details of the mechanism confining 
standard model particles and four-dimensional modes of gravity on the brane).
This introduces two alternative scenarios:
\par
\noindent
i) for $q\gg 1$, ``lightweight black holes'' (LBH) with $\mc\lesssim M\lesssim \mg$ may exist
and be produced at the LHC even if $\mg\gtrsim 10\,$TeV.
In this case, the effects due to bulk graviton exchanges would remain undetected;
\par
\noindent
ii) if $q\ll 1$, black holes do not exist with $M\simeq \mg$, even if $\mg\simeq 1\,$TeV,
and processes involving bulk gravitons are the only available signature
of extra-spatial dimensions.
\par
The former scenario might have important phenomenological implications both for
accelerator physics and in astrophysics.
Although recent LHC data at $7\,$TeV center mass energy seem to exclude the
production of microscopic black holes~\cite{cms,atlas}, there is still the possibility that
future runs at $14\,$TeV will achieve this goal.
Further, LHB might play a role in cosmological models as primordial black holes
produced in the early universe, and in astrophysics as the outcome of high energy
cosmic rays colliding against dense stars~\cite{astro}.  
\par
The case of $q\ll 1$ might instead explain why there is no evidence of black holes
at the LHC~\cite{cms,atlas}, even if extra-spatial dimensions exist.
\acknowledgments
We would like to thank B.~Harms and J.~Ovalle for stimulating discussions and suggestions,
X.~Calmet, C.~Kiefer, G.~Landsberg and E.~Winstanley for comments.
G.L.~A. and R.C.~thank Bologna ATLAS people, in particular L.~Bellagamba, G.~Bruni
and B.~Giacobbe, and CMS people, in particular A.~Perrotta.
Part of this work was supported by the COST~Action~MP0905 - Black Holes in a Violent Universe.
\end{document}